\documentclass[preprint,showkeys,aps,prb]{revtex4}
\usepackage[dvips]{graphicx}
\begin{document}

\title{                                                                                                     
The Simplest Viscous Flow
}

\author{
William Graham Hoover with Carol Griswold Hoover \\
Ruby Valley Research Institute                   \\
Highway Contract 60, Box 601                     \\
Ruby Valley, Nevada 89833                        \\
}

\date{\today}


\vspace{0.1cm}

\begin{abstract}
We illustrate an atomistic periodic two-dimensional stationary shear flow,
$u_x = \langle \ \dot x \ \rangle = \dot \epsilon y$, using the simplest possible example,
the periodic shear of just two particles ! We use a short-ranged ``realistic'' pair potential,
$\phi(r<2) = (2-r)^6 - 2(2-r)^3$.
Many body simulations with it are capable of modelling the gas, liquid, and solid states of
matter. A useful mechanics generating steady shear follows from a special (``Kewpie-Doll''
$\sim$ ``$qp$-Doll'') Hamiltonian based on the Hamiltonian coordinates $\{ q \}$ and momenta
$\{ p \}$ : ${\cal H}(q,p) \equiv K(p) + \Phi(q) + \dot \epsilon \sum qp$. Choosing $qp \rightarrow
yp_x$ the resulting motion equations are consistent with steadily shearing periodic boundaries
with a strain rate $(du_x/dy)  = \dot \epsilon$. The occasional $x$ coordinate jumps associated with
periodic boundary crossings in the $y$ direction provide a Hamiltonian that is a piecewise-continuous
function of time. A time-periodic isothermal steady state results when the Hamiltonian motion equations
are augmented with a continuously variable thermostat generalizing Shuichi Nos\'e's revolutionary ideas
from 1984. The resulting distributions of coordinates and momenta are interesting multifractals, with
surprising irreversible consequences from strictly time-reversible motion equations.

\end{abstract}

\maketitle

\pagebreak

\section{Introduction to Isothermal Molecular Dynamics}

Molecular Dynamics began in the 1950s with Fermi, Pasta, and Ulam's investigation of one-dimensional
anharmonic chains. That work was soon followed with Alder and Wainwright's dynamical exploration of
the pressure/volume equations of state for two-dimensional hard disks and three-dimensional hard spheres. While Fermi
was surprised that his chains never reached equilibrium, Alder and Wainwright found that disks and
spheres equilibrated rapidly, in just a few collision times. Alder and Wainwright also produced
evidence for first-order melting/freezing transitions for hard particles. These earliest molecular
dynamics simulations used from dozens to hundreds of particles, solving Newton's motion equations. By
2008 quadrillion-atom molecular dynamics $(N = 10^{12})$  was feasible\cite{b1}. Germann and Kadau
found that a single timestep, for a short-ranged pair potential in an unstable cubic lattice and with $N =
10^4 \times 10^4 \times 10^4$ particles, took about a minute of computer time to execute.

{\it Periodic} boundary conditions are the simplest choice, with any particle exiting at a system boundary
simultaneously introduced on the opposite side. In the absence of phase transitions such homogeneous
100-particle simulations are sufficient to determine the mechanical pressure/volume and thermal
energy/temperature equations of state with accuracies of order 1\%.

Nonequilibrium simulations require more sophisticated boundary conditions. In his 1974 thesis work at
Livermore Ashurst sandwiched rectangular Newtonian manybody systems between two ``fluid-wall''
regions.\cite{b2} The particles in these bounding walls had specified values of the mean velocity and
the mean-squared velocity imposed at every timestep. With different boundary velocities the resulting
shear viscosity could be measured. With different boundary kinetic temperatures,
$K/N \sim T \sim \langle \ (p^2) \ \rangle$,
heat flow could be simulated and the thermal conductivity measured. By 1980 it was well accepted that
the equilibrium and nonequilibrium properties for manybody systems provided excellent models for the
behavior of real fluids, both at, and away from, equilibrium.

In 1984 Shuichi Nos\'e formulated a temperature-dependent Hamiltonian reproducing Gibbs' canonical-ensemble
distribution of energies, $\propto \exp[-E/kT]$. His formulation\cite{b3,b4} included a relaxation time
$\tau$ with rapid relaxation consistent with Ashurst's rescaling of the mean-squared velocity to impose
temperatures within the fluid walls. The ``Nos\'e-Hoover'' version of this work\cite{b5} is better suited
to computation than was Nos\'e's original Hamiltonian-based algorithm. For $N$ particles of unit mass in $D$
space dimensions the equations of motion have the form:
$$
\dot q = p \ ; \ \dot p = F(q) - \zeta p \ ; \ \dot \zeta = [ \ (\sum p^2/NDkT) - 1 \ ]/\tau^2 \ .
$$
For $D=2$ $q$ and $p$ have both $x$ and $y$ components: $q = (x,y) \ ; \ p = (p_x,p_y)$.
$k$ is Boltzmann's constant (for convenience chosen equal to unity) and $T$ is the specified kinetic
temperature, $\langle \ p^2/2k \ \rangle = \langle \ p^2/2 \ \rangle .$

The ``friction coefficient'' $\zeta$ imposes integral control on the kinetic temperature in a time of order
$\tau$.  In the instantaneous limit, $\tau \rightarrow 0$, integral control corresponds to Ashurst's velocity
rescaling. In the equilibrium case, without external driving, Hoover pointed out that the time-averaged
distribution of the friction coefficient $\zeta$ is Gaussian, with zero mean. We will see that this changes
away from equilibrium where $\langle \ \zeta \ \rangle$ corresponds to entropy production, and is necessarily
positive.

\section{Imposing Shear on Molecular Dynamics with Doll's Tensor}

With controlled boundary conditions molecular dynamics can be generalized to {\it nonequilibrium} flows
characterized by viscosity and thermal conductivity.  These material properties characterize the response
of fluids or solids to gradients in the velocity or the temperature. Viscosity is the simpler of these
possibilities. Periodic boundary conditions are natural for shear deformation.

A straightforward viscous modification of equilibrium molecular dynamics can be implemented by imparting
a systematic motion to the neighbors of a periodically-repeated simulation, as illustrated in {\bf Figure 1}.
The figure shows the details of two particles interacting periodically. Three parallel periodic rows,
of $3N=6$ particles each are required. There is a systematic boundary
motion in the $x$ direction which increases linearly with $y$. The ``Doll's Tensor'' Hamiltonian (named after
the $qp$ term reminiscent of the Kewpie Doll) consistent with this motion was introduced by Hoover in
1980\cite{b6}. With $(qp) \rightarrow (yp_x)$ this Hamiltonian is
$$
{\cal H}_{\rm Doll's} = \sum (p^2/2) + \Phi + \dot \epsilon \sum yp_x \rightarrow
\{ \ \dot x = +(\partial{\cal H}/\partial p_x) = p_x + \dot \epsilon y \ \} \ .
$$
The new term proportional to the strain rate, $\dot \epsilon =(du_x/dy)$, along with periodic boundaries in
the $x$ and $y$ directions, generates a shear flow.  With just two particles, such a model provides the simplest
atomistic simulation of viscous shear flow. The Doll's Tensor equations of motion, applied to momenta $(p_x,p_y)$,
with vanishing sums, produce a systematic increase of $x$ velocity with $y$, $\dot \epsilon y$:
$$
\{ \ \dot x = p_x + \dot \epsilon y \ ; \ \dot y = p_y \ ; \
\dot p_x = F_x \ ; \ \dot p_y = F_y - \dot \epsilon p_x \ \} \ .
$$
For simplicity we consider forces $\{ \ F_x,F_y \ \}$ derived from a pair potential $\phi(r)$:
$F(r) = -\nabla \phi(r)$.
In this case the total potential is a simple sum over all pairs of particles, $\Phi = \sum \phi_{i<j}$.

With pair forces the change in the ``internal'' energy $E = \sum(p^2/2) + \Phi$
is precisely consistent with the first-law thermodynamic relation for energy conservation in the
absence of heat flow, $\Delta{\rm Heat} = 0 \longrightarrow \Delta E = \Delta {\rm Heat} - \Delta {\rm Work} = - \Delta{\rm Work}$:
$$
\dot E = -\dot \epsilon \sum^N(p_xp_y) -\dot \epsilon \sum^{\rm pairs} (Fxy/r)_{i<j} =
-\dot \epsilon P_{xy}V \ .
$$
Here the force-dependent pair sum includes the particle separations in the $x$ and $y$ directions:
$$
(xy/r) \equiv (x_i-x_j)(y_i-y_j)/\sqrt{[(x_i-x_j)^2+(y_i-y_j)^2]} \ .
$$
Notice particularly that the First Law of Thermodynamics just given is perfectly
time-reversible. In a reversed process $\dot E$ necessarily changes sign, as does
the strain-rate $\dot \epsilon$. The shear component of the pressure tensor does not
change with the direction of time, thus violating the empirical experience that
shearing a viscous fluid requires work, whether the shearing is clockwise or
counterclockwise. It is definitely paradoxical that the time-reversible laws of
microscopic mechanics provide simulations obeying the Second Law, that work can be
converted to heat, but not {\it vice versa}. We will extend this paradox, already
present in Hamiltonian mechanics, to a nonequilibrium mechanics in which both heat
and work enter, and with the same paradoxical disagreement with the Second Law.

We adopt a particularly simple pair potential, with a range $r<2$, and sufficiently
realistic to generate gas, liquid, and solid phases\cite{b7}. See {\bf Figure 2} for
a plot. This pair potential function is a simple polynomial in $r$, with an attractive
minimum of unity at a separation of unity, $\phi(r=1) = -1$:
$$
\phi_3^L(r) = (2-r)^6 - 2(2-r)^3 \rightarrow F = 6[(2-r)^5-(2-r)^2] \ .
$$
Thus the potential minimum sets both the length and the energy scales.  Both the force and
its first derivative vanish at the cutoff radius $r=2$, optimizing the numerical accuracy
of computer simulations.

The simplest viscous flow, illustrated in {\bf Figure 1}, is a two-particle dynamical
system. With a potential range of 2 a periodic $4 \times 4$ box is the smallest for
which the interaction of the two particles involves just one of the 9 possibilities
illustrated in the figure. With vanishing center-of-mass velocity the two-particle 
problem is equivalent to a one-particle one, with the boundary conditions including the
steady motion of cells containing image particles. In the two-particle version the four
time-dependent variables $(x,y,p_x,p_y)$ for Particle 2 exactly equal the negatives of
the same variable set for Particle 1.

Let us turn next to the integration details. We like to avoid the need for developing
special integration algorithms for the various types of problems described by atomistic
simulations. For simplicity and accuracy we prefer fourth-order  Runge-Kutta integration
for all those problems that are describable with first-order differential equations\cite{b8}.
These included models in which constraints on the momentum or temperature or stress
are imposed by Hamiltonian or nonHamiltonian constraints. Let us review our favorite
Runge-Kutta integrator by applying it to two instructive problems.  The first is the
simple harmonic oscillator. This problem is useful for code development as the
Runge-Kutta algorithm has an explicit analytic solution for the oscillator.

We introduce additional complexity and nonlinearity in our second pedagogical example,
the periodic dynamics of two two-dimensional particles using the realistic pair potential
just described, a finite-ranged very smooth polynomial in the separation. The boundaries
for that problem simplify in the absence of the systematic velocity gradient of {\bf Figure 1}.
The static-boundary  problem appears to be chaotic and might even be ergodic despite
its simplicity. The problem's description can be reduced from eight dimensions to just three
by taking the constancy of the center of mass, its motion, and the conserved total energy into
account. Three dimensions is just sufficient for chaos. Let us consider the details of the
Runge-Kutta integration algorithm and its applications to the harmonic oscillator and the
two-dimensional periodic two-body problem next.

\pagebreak

\section{Two Problems Illustrating the Runge-Kutta Algorithm}

\subsection{The Fourth-Order Runge-Kutta Integration Algorithm}
For completeness we review the programming necessary to implement the fourth-order
Runge-Kutta integrator. The simplicity and accuracy of this integrator make it our first
choice whenever possible. Consider solving $N$ first-order differential equations using
the Runge-Kutta algorithm.

A main program calls the fourth-order Runge-Kutta subroutine to execute each timestep,
advancing the $N$, ${\tt N}$ in the subroutine, time-dependent variables, stored in the
vector ${\tt YNOW}$. In the subroutine, given below, the righthandsides of the $N$
first-order differential equations are evaluated successively for four different vectors,
providing an averaged time derivative which is accurate through the fourth power of the
timestep $dt$, {\tt DT} in the program. As the timestep is specified at the beginning of
each step, it can be changed ``on the fly'', if necessary.  For instance, a useful
diagnostic for accuracy is the difference between the result of a single step $dt$ and
that of two successive steps half as large, $dt/2$. If the difference is too large the
step {\tt DT} can be reduced to ${\tt DT/2}$. If the difference is too small ${\tt DT}$
can be doubled.  The fourth-order Runge-Kutta algorithm is called at each
timestep and has the effect of converting the $N$-dimensional solution vector {\tt YNOW(T)}
at time $t$ to its value at time $t+dt$: ${\tt YNOW(T)} \rightarrow {\tt YNOW(T+DT)}$.

\pagebreak

\begin{verbatim}
  CALL RHS(YNOW,YDOT)
  DO 1 J = 1,N
1 D1(J) = YDOT(J)
  DO 2 J = 1,N
2 YNEW(J) = YNOW(J) + DT*D1(J)/2
  CALL RHS(YNEW,YDOT)
  DO 3 J = 1,N
3 D2(J) = YDOT(J)
  DO 4 J = 1,N
4 YNEW(J) = YNOW(J) + DT*D2(J)/2
  CALL RHS(YNEW,YDOT)
  DO 5 J = 1,N
5 D3(J) = YDOT(J)
  DO 6 J = 1,N
6 YNEW(J) = YNOW(J) + DT*D3(J)
  CALL RHS(YNEW,YDOT)
  DO 7 J = 1,N
7 D4(J) = YDOT(J)
  DO 8 J = 1,N
8 YNOW(J) = YNOW(J) + DT*(D1(J)+2*(D2(J)+D3(J))+D4(J))/6
\end{verbatim}

\subsection{Hamiltonian Dynamics of the Harmonic Oscillator}

The simplest instructive application of the Runge-Kutta integrator is the harmonic
oscillator example. Only two motion equations are involved, $\dot q = +p$ and
$\dot p = -q$. With $(q,p)$ initially $(+1,0)$ the analytic solution has a period of
$2\pi$:  $q = \cos(t) \ ; \ p = -\sin(t)$.  Dividing the interval $0 < t < 2\pi$ into
100 steps with initial values {\tt YNOW(1) = 1, YNOW(2) = 0} gives a double-precision
estimate of the algorithm's accuracy. The error in $q^2 +  p^2$ at $t=2\pi$ (twice the
final total oscillator energy) is $8.5415 \times 10^{-8}$. The history of the decrease
is accurately linear in time, as is illustrated in {\bf Figure 3}. Reducing the timestep
twofold gives an error nearly 32 times smaller, $2.6702  \times 10^{-9}$. The error ratio,
31.99, confirms that the energy error varies as $dt^5$, one order better than the larger
seperate errors in the coordinate and momentum.

Notice that the input vector {\tt YNOW} is incremented by the mean value of four estimated
{\tt RHS} vectors in each Runge-Kutta iteration. The derivatives {\tt D1, D2, D3, D4} are
evaluated at the initial point and three subsequent nearby points.  When the righthandsides
of the differential equations are explicitly time-dependent, as in periodic shear, the four
successive evaluations of the righthandsides in each timestep are evaluated at times {\tt T}
for {\tt D1}, {\tt T + DT/2} for {\tt D2} and {\tt D3}, and finally, {\tt T + DT} for
{\tt D4}.

The fourth-order Runge-Kutta solution of the oscillator equations $\dot q = p$ and $\dot p
= -q$ with $[q,p]$ initially $[+1,0]$, has a relatively simple and useful  analytic form:
$$
[q_n,p_n] = [1 - (dt^6/72)+(dt^8/576)]^{n/2}[+\cos(n\lambda),-\sin(n\lambda)] \ {\rm with}
$$
$$
\lambda \equiv \arctan \frac{[dt - (dt^3/6)]}{[1-(dt^2/2)+(dt^4/24)]} \ .
$$
For sufficiently {\it large} $dt$, so that $dt^8/576$ is properly represented, this result
provides a useful and informative programming check for the implementation of
the Runge-Kutta integration algorithm. We can see that $(q^2 + p^2)$, rather than being
constant at long times $ndt$, decreases as $ndt^6/72 = tdt^5/72$, one order of $dt$ better
than would be expected for a fourth-order method. Although the oscillator amplitude is
fifth-order, both the coordinate and the momentum $[q,p]$ do exhibit errors of order $tdt^4$.
Further analysis shows that this difference in orders is due to a ``phase error'' with the
oscillator slowed to fourth order in $dt$, rather than reduced in amplitude. Thus the dominant
error at a fixed time is a negative phase error of order $tdt^4$. With twenty timesteps per
period a double-precision comparison of the numerical and analytical results from Runge-Kutta
up to a time $2\pi$ gives agreement to a dozen significant figures.

With this interesting oscillator task successfully completed, confirming the utility and
accuracy of the algorithm, fully-fledged nonlinear simulations with realistic interparticle
interactions as well as both static and dynamic  boundary conditions on the coordinates provide
the next steps toward our goal of simulating ``The simplest viscous flow''.

\pagebreak

\subsection{Equilibrium Dynamics with a Realistic Potential}

The equilibrium Runge-Kutta integration of two-body dynamics with static periodic boundaries
and with interparticle forces, $F(r<2) = - (d\phi/dr) = 6(2-r)^5 - 6(2-r)^2$, is an excellent
computational next step. Conservation of the energy, $E = \Phi + K = \phi_{12}+(p_1^2+p_2^2)/2$
is a (nearly) foolproof check on the programming.  To apply fourth-order Runge-Kutta integration
the eight time-dependent variables are stored in an eight-dimensional $yy$ vector.  For the
coordinates we choose
$$
yy(1) = x_1 \ ; \ yy(2) = x_2 \ ; \ yy(3) = y_1 \ ; \ yy(4) = y_2 
$$
and for the momenta
$$
yy(5) = p_{x_1} \ ; \ yy(6) = p_{x_2} \ ; \ yy(7) = p_{y_1} \ ; \ yy(8) = p_{y_2} \ .
$$

The $x$ coordinates are confined to the $4 \times 4$ periodic box, both initially and after
every integration step,  with four statements
\begin{verbatim}
if(x1.lt.-2) yy(1) = yy(1) + 4
if(x1.gt.+2) yy(1) = yy(1) - 4
if(x2.lt.-2) yy(2) = yy(2) + 4
if(x2.gt.+2) yy(2) = yy(2) - 4
\end{verbatim}
and likewise for the $y$ coordinates, $y1 = yy(3)$ and $y2 = yy(4)$.

The eight motion equations for the two-body problem are the same for both particles:
$$
\{ \ \dot x = p_x \ ; \ \dot y  = p_y \ ; \ \dot p_x = F_x \ ; \ \dot p_y = F_y \ \} \ ,
$$
To compute the forces the relative displacement of Particle 2 relative to Particle 1 is
first taken as the simple difference with both particles in the central box.
\begin{verbatim}
x12 = x1 - x2
y12 = y1 - y2
\end{verbatim}
Then the {\it nearest}-image distance between the two particles, smallest among the nine
possibilities visible in {\bf Figure 1}, is calculated by imposing two conditions on the $x$
separation,
\begin{verbatim}
if(x12.lt.-2) x12 = x12 + 4
if(x12.gt.+2) x12 = x12 - 4
\end{verbatim}
and likewise for ${\tt y12}$. It is convenient, and perfectly permissible, to impose the fixed
center-of-mass symmetry at the conclusion of each step: {\tt yy(J) = -yy(J-1)} for {\tt J}
equal to 2, 4, 6, and 8.

A bit of experimentation shows that a timestep $dt = 0.005$ provides good energy conservation.
At a time of 1000 the initial energy of unity has decayed to 0.9999937.  This time-dependent
energy decay is illustrated in {\bf Figure 3}, at the left for the oscillator and at the right
for the two-body problem. The two are also compared on a logarithmic scale in {\bf Figure 4}
with similar shapes, though the oscillator results are certainly smoother.

The jumps in the two-body energy, visible in both figures, occur at the collisional turning points.
In spite of the relative simplicity of this problem the long-time two-body trajectories illustrated in
{\bf Figure 5} suggest that there is sufficient chaos in this problem to provide a near-ergodic
coordinate distribution. Although the apparent phase-space dimensionality here is eight, much
more than the three required for chaos, this equilibrium problem is actually just three-dimensional !
The two-body problem, with the center-of-mass fixed, or equivalently, expressed in relative
coordinates, can be reduced to a four-dimensional problem. It could be viewed as the constant-energy
motion of a mass-point particle within a periodic $4 \times 4$ square with four fixed particles
at the corners controlling the mass point's motion. Conservation of energy further restricts that
motion, however expressed, to a three-dimensional energy shell, the actual minimum for chaos. In
fact isoenergetic simulations with unit energy show no tendency to approach either of the two
collisionless nonchaotic solutions:
$$
\{ \  x = \pm t \ ; \ y = \pm 1 \ ; \ p_x = \pm 1 \ ; \ p_y =     0 \ \} \ ;
\{ \  x = \pm 1 \ ; \ y = \pm t \ ; \ p_x =     0 \ ; \ p_y = \pm 1 \ \} \ .
$$
In these hypothetical collisionless cases the motion would parallel either of the two coordinate
axes, $x$ or $y$.

The chaotic equilibrium problem, just described and pictured in {\bf Figure 5} is a useful
stepping stone toward the stationary two-body nonequilibrium shear flow that is our goal.  Let us
consider next the modelling of a steadily moving periodic boundary. Incorporating this driving
mechanism will help us simulate a ``stationary'', actually time-periodic, nonequilibrium flow. The
moving boundary introduces a new variable, the {\it phase} or ``strain'' of the motion, which
can be reset to zero when the coordinates in all nine cells regain their perfect square-lattice
symmetry.

There is a second barrier to stationarity. Because the moving boundaries do viscous
``work'' on the two particles in the central cell we will need to ``thermostat'' the motion to
avoid a relatively rapid divergence of the kinetic energy.  We will use a Nos\'e-Hoover thermostat
force, $-\zeta p$ to control the kinetic energy. With the addition of a moving boundary and a
thermostat our two-body equilibrium model is converted to a nonequilibrium steady flow.  We
describe its characteristics next. For simplicity we restrict our description to the case of
unit strain rate so that the locations of the periodic-image cells repeat at integral values of the
time.

\section{Nonequilibrium Shear Flow with Two Particles}

To reach our goal of ``The Simplest Viscous Flow'' requires two additions to the equations of motion
and corresponding modifications of the integrator. Applying the Runge-Kutta algorithm to shear
flow requires augmenting the motion equations to include [1] the explicit time dependence of the
boundary conditions and [2] the control of the kinetic energy. Let us consider the modifications in
more detail.

\subsection{Imposing a Fixed Strain Rate on the Dynamics}

In introducing shear the three square cells with $-2 < y < +2$ obey the usual periodic boundary
conditions in the $x$ direction. The $y$ direction is different.  The shearing motion of three
square cells above and three below the central cells is induced by giving the twelve particles
in the upper and lower cells additional velocities equal to the product of the strain rate
and the cell height: $\dot x = p_x +4$ above, $\dot x = p_x - 4$ below. We choose to restrict the
cell-to-cell strain [ $\epsilon = (\Delta x/\Delta y)$ relative to the perfect square-lattice
structure ] of the
nine cells of {\bf Figure 1} to lie in the range from $-0.5 < \epsilon < +0.5$. From the numerical
standpoint simply resetting the strain, $+(1/2) \rightarrow -(1/2)$, every 200 steps, prevents the
loss of significant figures in the horizontal coordinates $x$. Otherwise simulations with billions
of timesteps, corresponding to strains on the order of millions, would erode the accuracy of the
horizontal coordinates.

By limiting the strain to lie between the two extreme values $\mp(1/2)$ shown in the figure, it
is guaranteed that the nearest-neighbor image of Particle 1 lies uniquely in one of the nine cells
shown. Likewise for Particle 2. Thus the nearest-image potential and forces can both be computed
for the central-cell Particles 1 and 2 by selecting the minimum-distance particle pair out of the
nine possibilities. The unique pair $(x_{ij},y_{ij})$ which minimizes the separation
$\sqrt{(x_{ij}^2 + y_{ij}^2)}$ is selected from the nine possibilities whenever the righthandsides
of the differential equations are calculated, four times within every Runge-Kutta timestep.

\subsection{Imposing a Fixed Time-Averaged Temperature on the Dynamics}

We choose a basic time-reversible Nos\'e-Hoover thermostat to control the temperature. In the
time-reversed version both the momenta and the friction coefficient $\zeta$ change signs. For
simplicity we choose a thermostat relaxation time of unity in the $\dot \zeta$ equation. The
thermostatted motion equations are the following, where $T$ is a specified constant target
temperature:
$$
\{ \ \dot x = p_x + \dot \epsilon y \ ; \ \dot y = p_y \ ; \
\dot p_x = F_x - \zeta p_x\ ; \ \dot p_y = F_y - \dot \epsilon p_x - \zeta p_y \ \} \ ; \
$$
$$
\dot \zeta = K - NT \ {\rm where} \  K = \sum(p_x^2 + p_y^2)/2 \ .
$$
These equations are time-reversible. To see this note first that the overdot ``$ \ ^{\bf .} \ $''
signifies a
time derivative so that ($\dot x,\dot y$) change signs with the direction of time while ($\dot p_x,
\dot p_y$) and the forces do not. Evidently $\dot \epsilon$ must change sign for reversibility, in
agreement with our imagining the shear process from $-0.5$ to $+0.5$ being reversed, from $+0.5$
to $-0.5$. Likewise the sign of the friction coefficient must be reversed just as if it were a
momentum.  In fact, in Nos\'e's original derivation of his thermostat equations the friction
coefficient {\it was} a momentum. Its conjugate variable $s$ is Nos\'e's ``time-scaling variable'', which
behaved like a coordinate.  Though the motion equations just given are patently time-reversible we
will see, and come to understand, that the reversibility is illusory, and if attempted will succeed
for only a relatively short time.

In this shear flow problem there is a competition between the driving shear force, characterized
by the strainrate $\dot \epsilon$, and the damping activity of the thermostat, characterized by
$\zeta$. These two forces represent work and heat respectively. $\epsilon$ provides work and the
thermostat variable $\zeta$ on average extracts the resulting irreversible heating, maintaining a
kinetic temperature $2T = \langle \ K \ \rangle = \langle \ (p_1^2+p_2^2)/2 \ \rangle $. With a
strainrate of unity a short computation shows that a thermostated kinetic energy of 0.3 provides a
longtime chaotic solution.

At lower temperatures an interesting phenomenon ensues. Low temperature favors an interparticle
spacing of unity, the energy minimum. {\bf Figure 6} displays exactly this. At lower temperatures,
with kinetic energies in the range from 0.25 down to zero, the energy minimum becomes irresistable
and the fractal distributions, like
those generated at 0.30 and higher, collapse to a fixed point in the nine-dimensional phase space.
Such points correspond to an interparticle separation barely exceeding unity, and close to the
potential minimum for the interacting Particles 1 and 2. The two particles occupy fixed points on a
curve approximating a circle, with diameter slightly greater than unity.

As a consequence, once sufficiently cooled, Particles 1 and 2 no longer can interact with their
periodic images to the sides or above or below, for which a central-box separation of the particles
greater than 2 is required. Thus low-temperature simulations are no longer representative of boundary-driven
shear. {\bf Figure 6} shows these unphysical fixed-point solutions. With this thermostat model
the ``temperature'' is entirely taken up by $p_x^2$ and exactly cancels the coordinate-dependent
contribution $\dot \epsilon y$ to $\dot x$.  Meanwhile $p_y$ and $\dot y$ vanish.
$$
\dot x = p_x + \dot \epsilon \times y = 0.5 + (1.0 \times -0.5) = 0 \ (\rm{For} \  2T = K = 0.25 \
{\rm and} \ \dot \epsilon = 1) \ .
$$
Because our pair potential has its maximum value $2^6-2^4=48$ at $r=0$ it is likewise evident that
coupling with the boundary will have a much diminished effect at sufficiently high temperatures. The
high-temperature limit is simply uniform over the $4 \times 4$ square, assuming sufficient
perturbation from the thermostat to favor irrational coordinates and momenta over the rational.
Chaotic viscous flow can only exist when both the driving and the dissipative influences
on the motion are comparable, so that their time-averaged contributions cancel.

{\bf Figure 7} pictures an $(x,y)$ projection of the fractal distribution at $K = 2T = 0.30$.
In the figure the six boundary-cell images of Particles 1 and 2 in the cells above and below the central
cell are swept from left to right (above) and right to left (below) accounting for the smearing of
probability at the top and bottom of the central cell. {\bf Figure 8} shows superpositions of snapshots
taken every 200 steps (corresponding to a unit increase in $\epsilon = \dot \epsilon t$). Just two of
the 200 different phases, at integral and half-integral strains respectively, making up a unit time
interval, are shown in the figure. Evidently the probability density is periodic (with period 200
steps here) in time so that a detailed microstate description would need to include the phase of the
motion as well as the the values of the coordinates and momenta of Particles 1 and 2.

{\bf Figure 9} shows two phases of the two-body momentum distribution. The kinetic part of the pressure
tensor $P_{xy}$ is proportional to the summed values of $p_xp_y$, identical for the two particles in
our model.  The figure shows that the product is, on average, negative, as is required for agreement
with continuum mechanics where the shear stress, the {\it negative} of $P_{xy}$, is equal to the product
of the strain rate and the (positive) coefficient of shear viscosity.

\section{Describing the Results of The Nonequilibrium Simulations}

\subsection{Paradoxical Nonequilibrium Phase-Volume Changes}

Let us consider the nonequilibrium shearing motion, time-averaged over many repetitions of the boundary
time-periodicity in five-dimensional space, $\{ x_1,y_1,p_{x_1},p_{y_1},\zeta \}$. Averaging the comoving
dilation rate in this phase space for an infinitesimal five-dimensional volume element $\otimes$ gives an
empirical inevitable inequality:
$$
\langle \ (\dot \otimes/\otimes) \ \rangle =
\langle \ (\partial \dot x_1/\partial x_1) + (\partial \dot y_1/\partial y_1) +
(\partial \dot p_{x_1}/\partial p_{x_1}) + (\partial \dot p_{y_1}/\partial p_{y_1}) +
(\partial \dot \zeta/\partial \zeta) \ \rangle =
$$
$$
 0 + 0 - \langle \ \zeta \ \rangle - \langle \ \zeta \ \rangle + 0 = - 2\langle \ \zeta \ \rangle < 0 \ .
$$
It is characteristic of time-reversible nonequilibrium stationary states that their phase volume
inevitably shrinks, as the time-averaged friction coefficient $\zeta$ is invariably positive.

From the computational standpoint a negative friction coefficient would correspond to comoving phase-space growth,
numerical instability, and numerical divergence. The inevitable positive friction coefficient, corresponding
to comoving shrinkage, and necessary for computational stability, at first looks paradoxical in view of the
formal time-reversibility of the motion. To understand, recall that ordinary equilibrium Hamiltonian mechanics
conserves the comoving phase volume precisely. This is a consequence of the Hamiltonian motion equations,
$$
\{ \ \dot q = +(\partial {\cal H}/\partial p) \ ; \ \dot p = -(\partial{\cal H}/\partial q) \ \} \
\longrightarrow
$$
$$
\sum [ \ (\partial \dot q/\partial q) + (\partial \dot p/\partial p) \ ] =
\sum [ \ (\partial/\partial q)(+(\partial{\cal H}/\partial p)) +
(\partial/\partial p)(-(\partial{\cal H}/\partial q)) \ ]
\equiv 0 \ . 
$$
{\it At} thermal equilibrium, imposed by Nos\'e-Hoover friction, the time-averaged friction coefficient
$\langle \ \zeta \ \rangle$ typically vanishes. Typically the equilibrium $\zeta$ has a Gaussian
distribution with zero mean. At equilibrium the logarithm of the phase volume $\otimes$ corresponds to entropy.
We will see that this changes away from equilibrium where $\langle \ \zeta \ \rangle$ corresponds to entropy
production, and is necessarily positive.

\subsection{A Microscopic Version of the Second Law of Thermodynamics}

Away from equilibrium, with driving forces and a compensating nonHamiltonian friction coefficient, the comoving
dilation rate $(\dot \otimes/\otimes)$ is nonzero. The internal energy $\Phi + K$ and the phase volume are no
longer constant. To
prevent the divergence of both energy and phase volume the mean friction coefficient is necessarily,
though paradoxically, positive. This positive friction, in the face of reversibility, is the microscopic
version of the Second Law of Thermodynamics, describing the conversion of work to heat. The reversed
version of this shearing process would be a violation of that law. This consonance of reversible microscopic
dynamics with macroscopic thermodynamics was at first surprising and controversial, but ultimately became both
edifying and satisfying, a major result obtained by analyzing the nonequilibrium simulations pioneered by Ashurst.
For two early time-reversible one-body ``work-to-heat'' simulations obeying the Second Law see References 9 and 10.

\subsection{The Nature of Nonequilibrium Lyapunov Spectra}

From the standpoint of dynamical systems the loss of phase volume is an expected feature for models generating
fractal distributions, but only rarely are these models time-reversible\cite{b11,b12}. An investigation of the Lyapunov
spectrum for fractal distributions invariably results in a negative sum, even in formally time-reversible
cases. The negative sum signals a collapse of phase volume to a zero-volume fractal, limit cycle, or fixed
point.

A worthy research goal is the Lyapunov analysis of the family of five-dimensional flow equations with 
ranges of strain rates $\{ \ \dot  \epsilon \ \}$
 and temperatures $\{ \ T \ \}$:
$$
  \{ \ \dot x = p_x + \dot \epsilon y \ ; \ \dot y = p_y \ ; \
\dot p_x = F_x - \zeta p_x\ ; \ \dot p_y = F_y - \dot \epsilon p_x - \zeta p_y \ \} \ ; \                            
$$
$$
\dot \zeta = (K/T) - 1 \ {\rm where} \  K = (p_x^2 + p_y^2)/2 \ .
$$
The averaged phase-space dilation rate, $\dot \otimes$, can be written in terms of the five long-time-averaged
Lyapunov exponents,
$$
\langle \ (\dot \otimes/\otimes) \ \rangle = -2\langle \ \zeta \ \rangle =
\langle \ \lambda_1 + \lambda_2 + \lambda_3 + \lambda_4 + \lambda_5 \ \rangle \ .
$$
Exactly this same equation holds instantaneously though the corresponding ``local'' values of the
exponents depend upon the chosen coordinate system.

The instantaneous values of the five Lyapunov exponents result from the analysis of growth rates
for five orthogonal time-dependent vectors in the space, $\{ \ \delta_1 \ \dots \ \delta_5 \ \}$.
Evidently the entire range of phase-space
dimensionalities from 0 (a fixed point) to 5 (ergodic coverage, which should result for reasonable
combinations of temperature with small strain rates) could be analyzed.  We heartily recommend
such studies to our readers. These simple dynamical problems are a relatively complex interesting
area ripe for study. Surely a periodic symmetric two-body problem is indeed the simplest viscous
flow.

\subsection{The Relevance of Time-Reversible Maps}

In 1996 Hoover, Kum, and Posch constructed and analyzed the results of several time-reversible
maps\cite{b13}. Those maps were all defined in a two-dimensional unit square with $-(1/2)<(x,y)<+(1/2)$.
Shears, for instance, obeyed the maps ${\bf X(x,y)}$, similar to the shearing considered here, and
${\bf Y(x,y)}$:
$$
{\bf X(x,y)}: x \rightarrow x^{\prime} = x + \epsilon y \ ; \ y \rightarrow y^{\prime} = y \ ;
$$
$$
{\bf Y(x,y)}: x \rightarrow x^{\prime} = x \ ; \ y \rightarrow y^{\prime} = y + \epsilon x \ .
$$
In these two-dimensional cartesian maps $x$ represents a coordinate and $y$ a momentum so that time
can be reversed by changing the sign of $y$, analogous to reversing time at a fixed configuration.
In this map modelling of time reversibility it is convenient to introduce the time-reversal map
${\bf T}$:
$$
{\bf T}(x,y): x \rightarrow x^{\prime} = x \ ; \ y \rightarrow y^{\prime} = -y \ . 
$$
With this definition ${\bf T}$ and its inverse ${\bf T^{-1}}$ are one and the same. In fact any
map ${\bf M}(x,y)$ is time-reversible if its inverse is the sequence of operations going
backward in time, ${\bf TMT}$, so that ${\bf MM^{-1}}(x,y)= {\bf M^{-1}M}(x,y) = (x,y)$.

The 1996 work showed that the mapping ${\bf XYYX}$ was ergodic, with constant density throughout
the unit square. Differently, and significantly, the mapping ${\bf XYPYX}$, where ${\bf P}$ was
both time-reversible and included regions of density expansion and contraction, produced {\it fractal}
distributions, ergodic for small deviations from the ${\bf XYYX}$ equilibrium and with an information
dimension reduced quadratically in the driving deviation away from equilibrium. Thus the two-dimensional
time-reversible maps already contain a mechanism for irreversibility resembling that found in the
present ``simple'' five-dimensional viscous flows.

\section{Acknowledgment}

We thank Clint Sprott, Karl Travis, and Kris Wojciechowski for their readings of a previous
draft. In particular Karl pointed out the importance of two-dimensional maps for the understanding
of the irreversibility of the present five-dimensional nonequilibrium flows. A particularly compelling
map is the compressible Baker Map, described in two arXiv articles, 1909.04526, ``Random Walk
Equivalence to the Compressible Baker Map and the Kaplan-Yorke Approximation to Its Information
Dimension'' and 1910.12642, ``2020 Ian Snook Prize Problem : Three Routes to the Information
Dimensions for a One-Dimensional Stochastic Random Walk and for an Equivalent Prototypical
Two-Dimensional Baker Map". As no suitable entries for the latter problem were received we again
offer a Snook-Prize cash award of one thousand United States dollars for the best work addressing
the 2020 problem received during the calendar year 2021. See our ``\$1000 SNOOK PRIZES FOR 2021:
The Information Dimensions of a Two-Dimensional Baker Map'', Computational Methods in Science and
Technology {\bf 21}, 93-94 (2021).

\section{Addendum of 8 July 2021}

Karl suggested that we include the second-order Runge-Kutta oscillator solution, with\\
\begin{center}
{\tt QNEW = Q + P*DT - Q*(DT*DT/2) ; PNEW = P - Q*DT - P*(DT*DT/2)} .\\
\end{center}
The solution is:
$$
[q_n,p_n] = [1 + (dt^4/4)]^{n/2}[+\cos(n\lambda),-\sin(n\lambda)] \ {\rm with}
$$
$$
\lambda \equiv \arctan \frac{dt}{[1-(dt^2/2)]} \ .
$$
Because this oscillator amplitude now increases exponentially in time the lack of molecular
dynamics simulations using second-order Runge-Kutta algorithms is unsurprising!

\pagebreak

\begin{figure}
\includegraphics[width=2.5 in,angle=-90.]{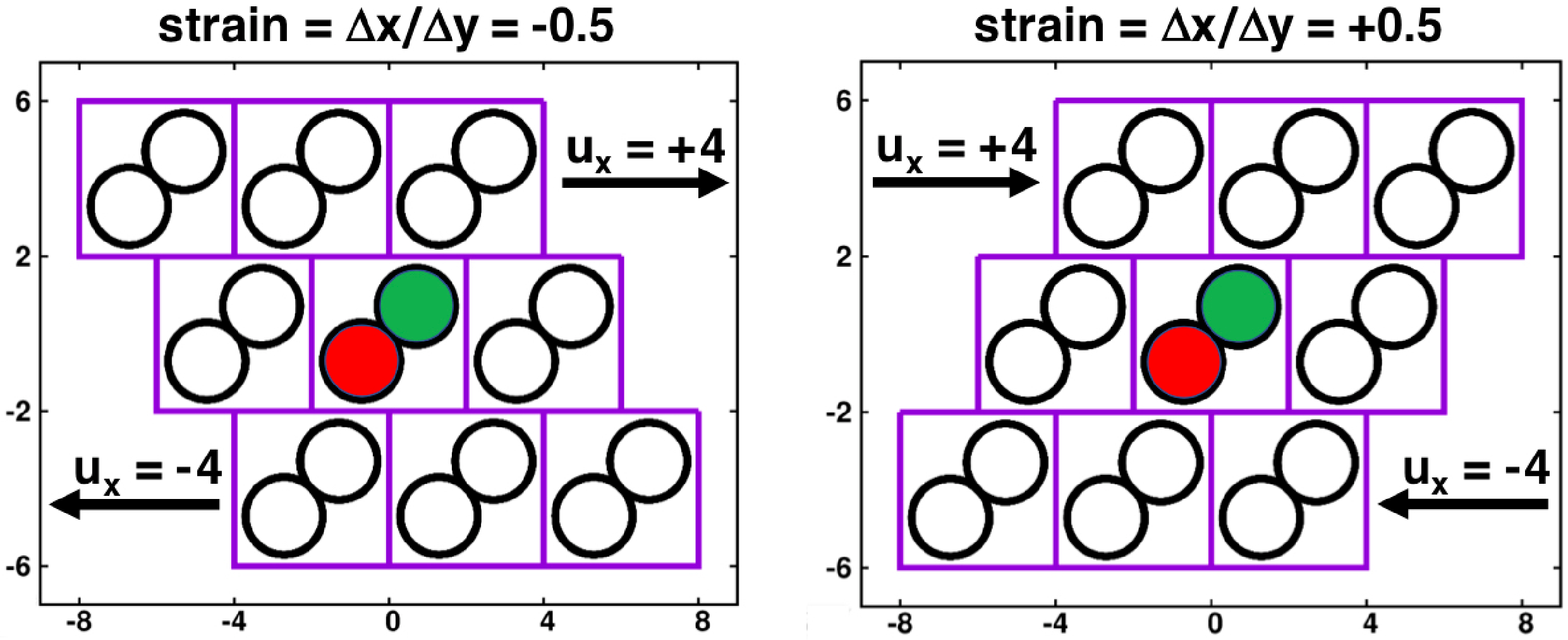}
\caption{
Steady time-dependent shearing boundary conditions are illustrated here for unit strainrate
$(du_x/dy)=1$ with a periodic $4\times 4$ cell containing Particles 1 and 2. The three upper
images of the central cell move to the right at $u_x = +4$ while the three lower images move
left with $u_x =-4$. Particles 1 and 2 (red and green) occupy the central cell with a fixed
center of mass at the origin.  The force on Particle 1 comes directly from Particle 2 or its
nearest image (from the eight neighboring cells) and is nonzero whenever that separation is
less than the range of the forces, $r_{12} < 2$. The force on Particle 2 is equal in magnitude
and opposite in sign, maintaining the center-of-mass position at the origin, the center of the
central cell. Whenever a horizontal cell boundary is crossed the systematic horizontal velocity
component $\dot x$ changes by $\pm 4$ and the $x$ coordinate undergoes a change equal to the
fractional part of the strain,  $\dot \epsilon t - [\dot \epsilon t]_{\rm integer}$, where $t$
is the time. The crossing of a vertical cell boundary changes only the value of the $x$
coordinate, by $\pm 4$, just as in the usual application of periodic boundary conditions. To
avoid the unnecessary loss of significant figures whenever the strain reaches $+0.5$, as shown
at the right, the strain is decreased by unity to $-0.5$, the strain shown at the left.
}
\end{figure}

\pagebreak

\begin{figure}
\includegraphics[width=5 in,angle= -0.]{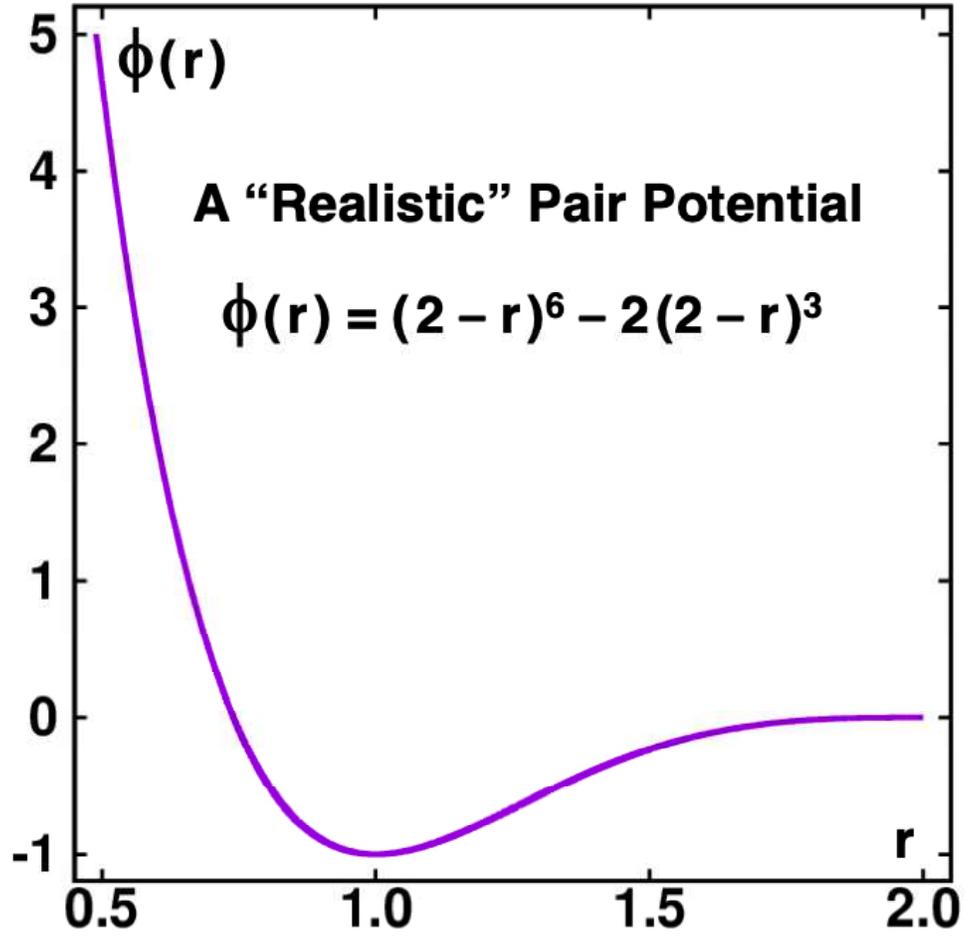}
\caption{
Pair potential for molecular dynamics simulation of shear, $\phi(r) = (2-r)^6 - 2(2-r)^3$. The
minimum at $r=1$ sets the energy scale and the distance scale. The vanishing of both the force
and its first derivative at $r=2$ provides highly-accurate Runge-Kutta trajectories.
}
\end{figure}

\pagebreak

\begin{figure}
\includegraphics[width=2. in,angle=-90.]{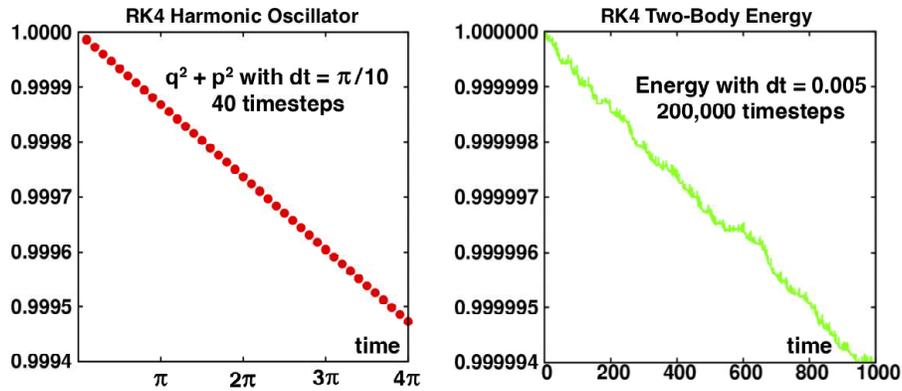}
\caption{
At the left we see the steady growth of energy error for the harmonic oscillator, with initial
conditions $(q,p) = (1,0)$, shown for forty time steps of $\pi/10$. The decay of the energy is
accurately linear in time with the error varying as $dt^5$ at a fixed time. The analytic
solution to this problem is given in Wm. G. Hoover's 1991 Elsevier book, {\it Computational
Statistical Mechanics}. At the right we see the analogous error growth for the Hamiltonian
dynamics of two two-dimensional particles with periodic boundary conditions for 200,000 steps
of 0.005 each. Though there are collisional fluctuations the growth rate is still roughly
linear and varies likewise as $dt^5$.
}
\end{figure}

\pagebreak

\begin{figure}
\includegraphics[width=5 in,angle=-0.]{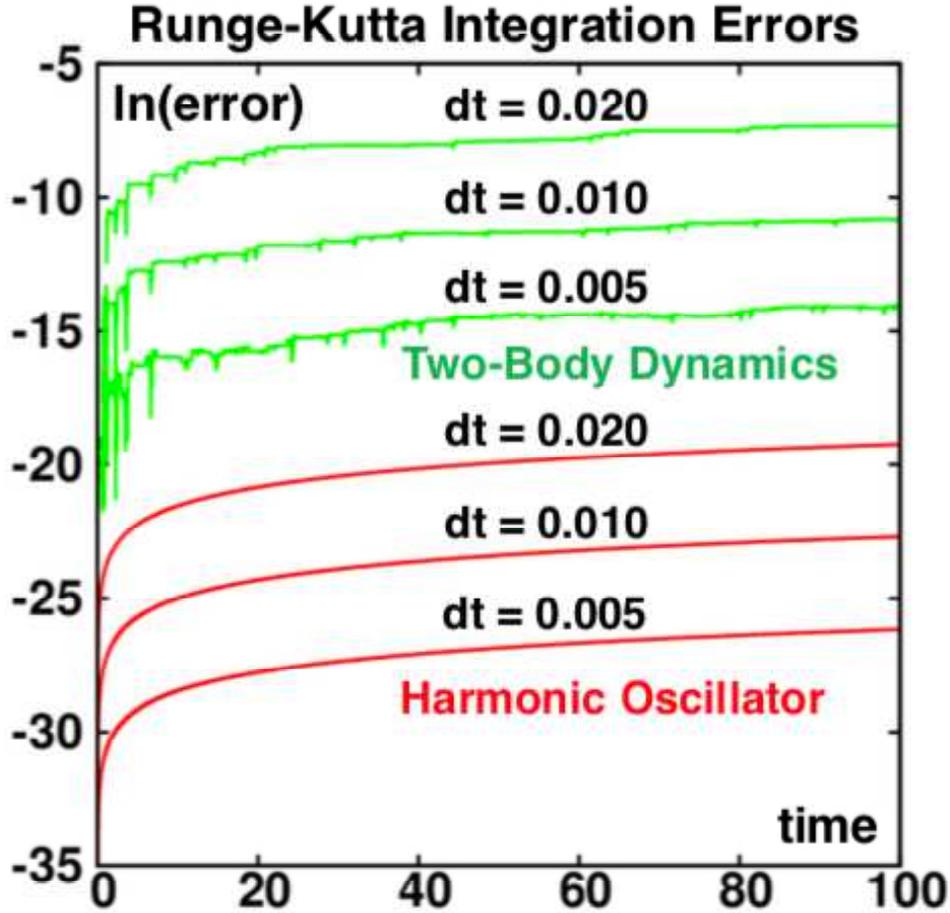}
\caption{
Comparison of the time-dependent error in $q^2+p^2$ for the harmonic oscillator (below) and the
periodic two-body equilibrium problem (above), with an initial energy equal to unity in the
latter case.  The analytic oscillator solution shows that the energy error at a fixed time varies
as $dt^5$, which the numerical work for two bodies nicely confirms.  Although the two-body
error increases less regularly, being dominated by the collisions, it is clear that the same
error analysis applies in that case. Replicating these results is a useful debugging exercise
for both these equilibrium Hamiltonian problems.
}
\end{figure}

\begin{figure}
\includegraphics[width=3.5 in,angle=-90.]{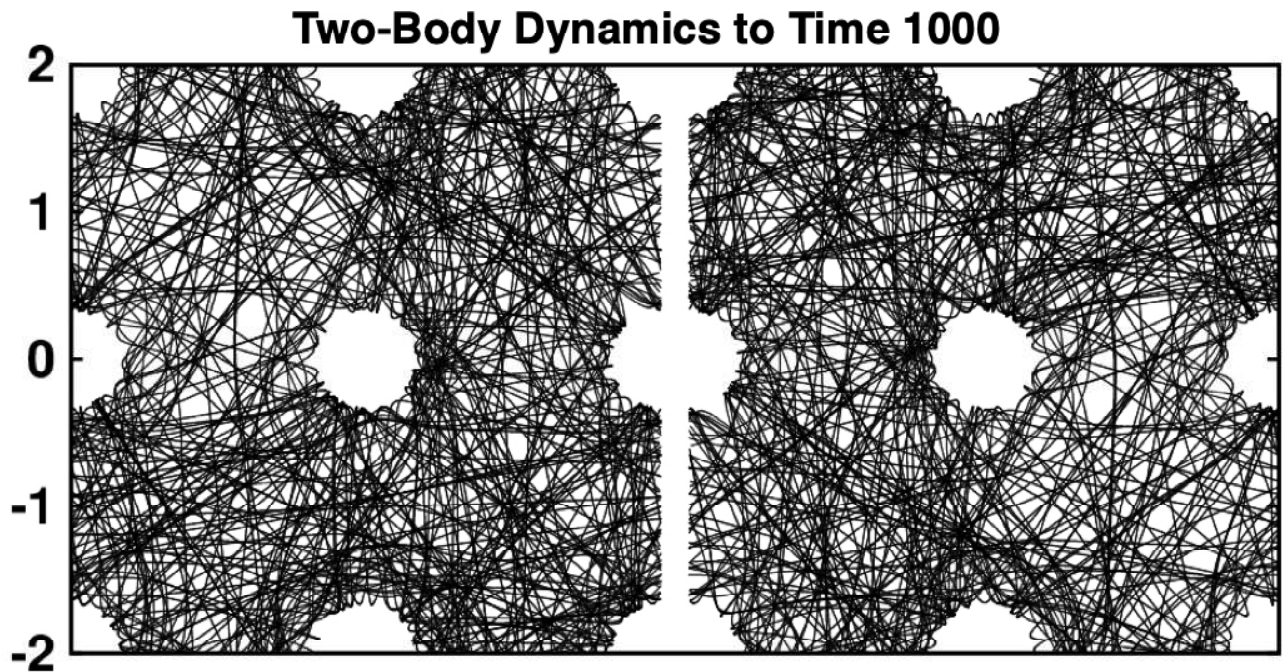}
\caption{
200,000 points along the equilibrium Hamiltonian $(x,y)$ trajectories for Particle 1 (at the
left) and Particle 2 (at the right). Notice the inversion symmetry resulting from the
center-of-mass constraints described in the text. The total energy of unity is conserved here.
The eight neighboring cells in this case are arranged in a fixed square lattice corresponding to
zero strain. Initially the two particles are separated as in {\bf Figure 1} with $r=2$,
$p_x = \mp 0.8$, and $p_y = \mp 0.6$, giving $E = 1$. The Runge-Kutta timestep is 0.005.
}
\end{figure}

\begin{figure}
\includegraphics[width=5.0 in,angle=-90.]{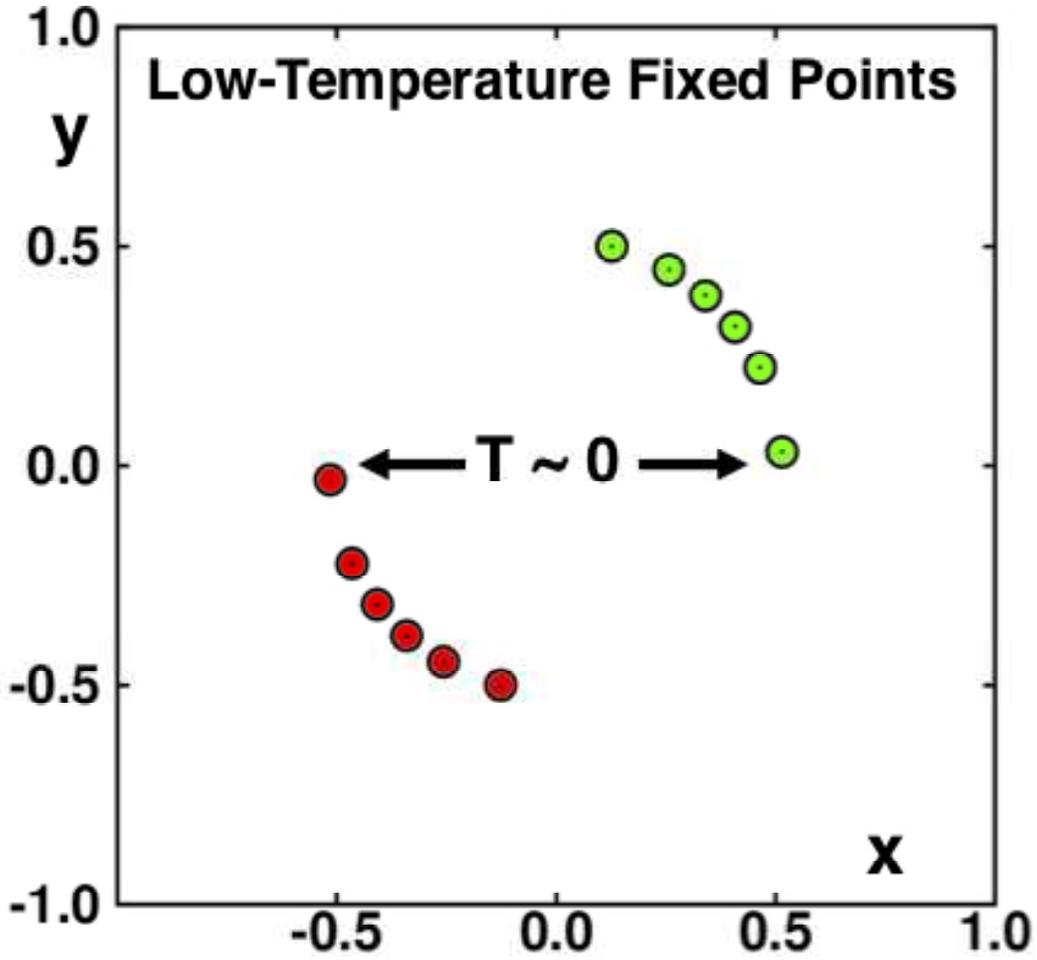}
\caption{
Particle locations for the fixed points at $p^2$-based kinetic energies from 0.00 to 0.25 in
steps of 0.05 with strainrate unity.  At a fixed point the horizontal motion vanishes so that
$\dot x = p_x + y = 0$. For Particle 1 the $y$ coordinate is $-\sqrt{K}$, precisely cancelling
the horizontal momentum. As the vertical momentum vanishes too $p_x^2 = K = 2T$. Convergence
at $T=0$ is poor while at kinetic energies from 0.05 to 0.25 convergence to machine accuracy
is rapid and accurate. The timestep is 0.005 and the friction coefficient obeys the motion
equation $\dot \zeta = K - 2T$.}
\end{figure}

\begin{figure}
\includegraphics[width=5. in,angle=-0.]{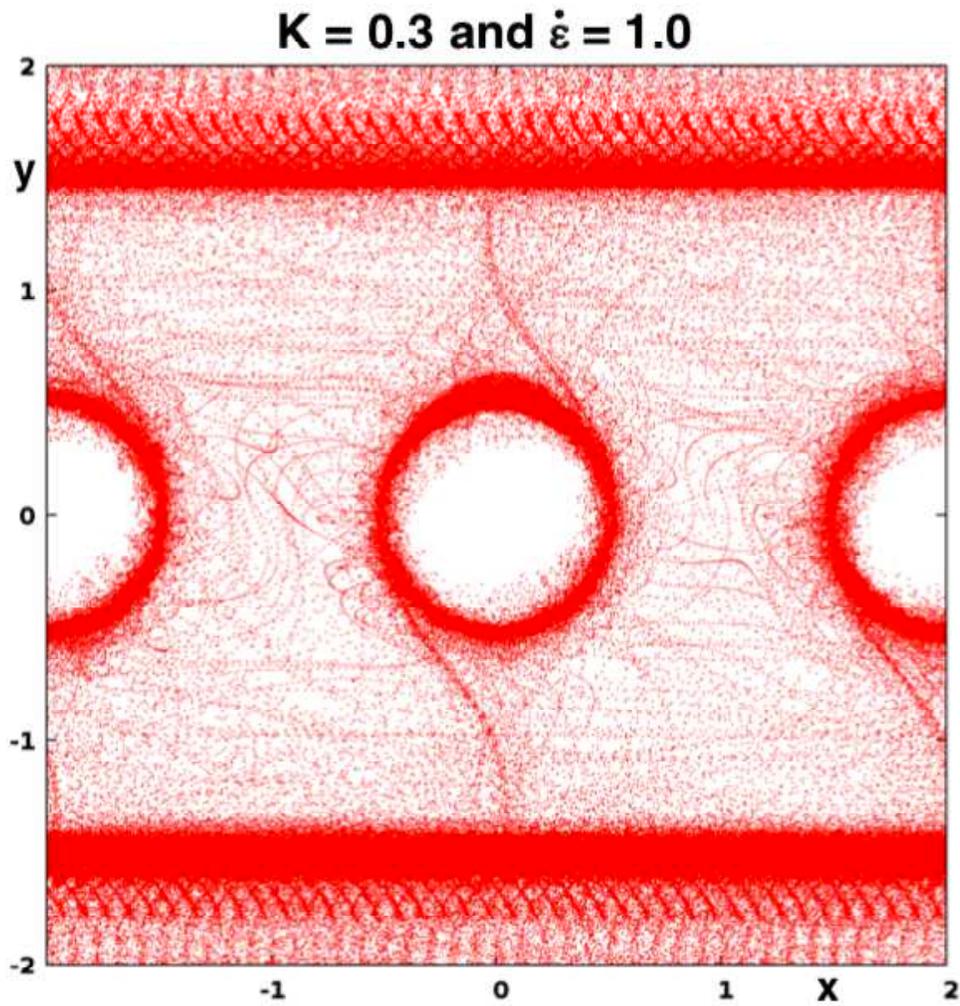}
\caption{
$(x_1,y_1)=(-x_2,-y_2)$ coordinates at unit shear strain rate with $\langle \ K \ \rangle = 0.3$
and $dt = 0.005$. The motion shown here covers a simulation from time 50,000 to time 100,000 as
represented by one million equally-spaced points.
}
\end{figure}

\pagebreak

\begin{figure}
\includegraphics[width=3. in,angle=-90.]{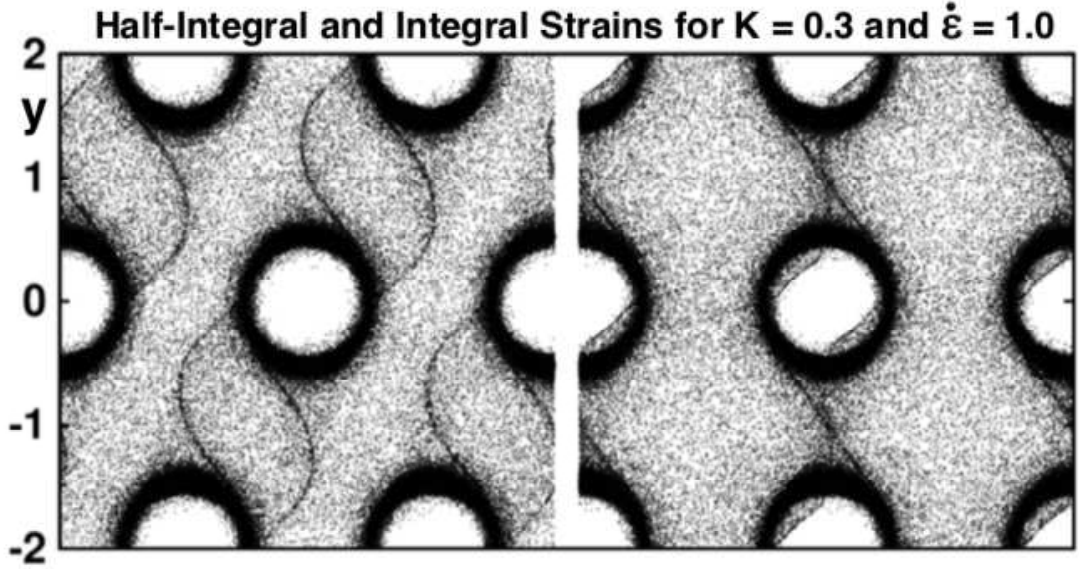}
\caption{
$(x_1,y_1)=(-x_2,-y_2)$ coordinates at integral (right) and half-integral (left) shear strains
from the last half of a 200,000,000 timestep simulation at unit strainrate and averaged kinetic
energy 0.3. One million points are shown.
}
\end{figure}

\pagebreak

\begin{figure}
\includegraphics[width=3. in,angle=-90.]{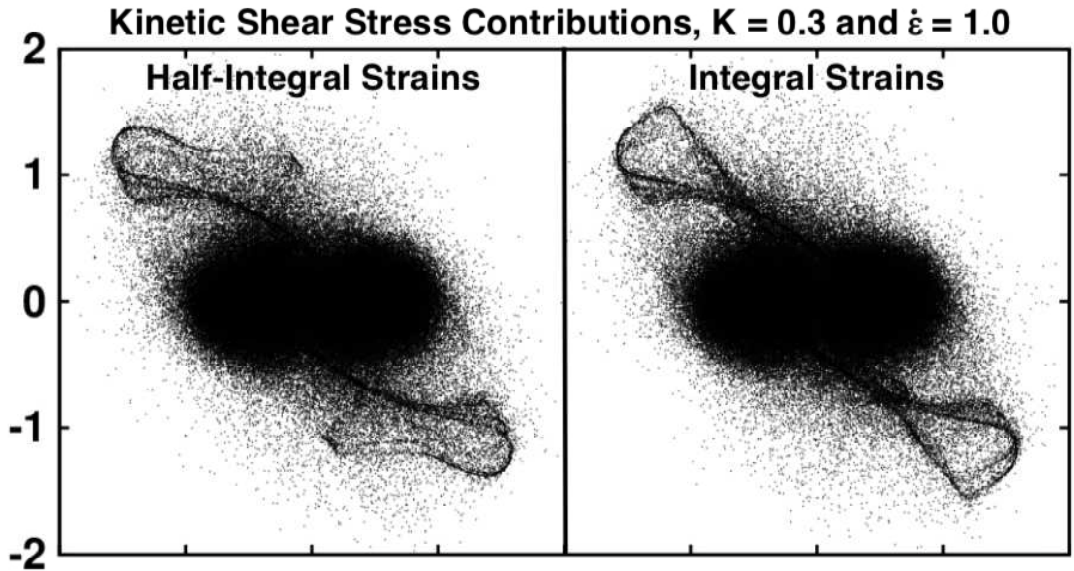}
\caption{
$(p_{x_1},p_{y_1})=(-p_{x_2},-p_{y_2})$ momenta at integral (right) and half-integral (left)
strains from shear strains beginning at $10^5$ and finishing at $10^6$. The timestep is 0.005,
the averaged kinetic energy $ \langle \ K \ \rangle = 0.3 = 2T$, and the shear strain rate
$\dot \epsilon=(du_x/dt)$ is unity. $\dot \zeta = K - 0.3$.  Evidently the velocity distribution
is time-periodic and sensitive to the phase of the boundary cells. Here 900,000 points are shown
in each panel, selected from the time interval from 100,000 to 1,000,000.
}
\end{figure}

\end{document}